\documentclass[12pt]{article}
\usepackage{amssymb}
\usepackage[dvips]{epsfig}
\usepackage[utf8]{inputenc}

\newcommand{\be}{\begin{equation}}
\newcommand{\ee}{\end{equation}}

\newcommand{\bn}{\begin{eqnarray}}
\newcommand{\en}{\end{eqnarray}}

\newcommand{\p}{\partial}

\newcommand{\nn}{\nonumber}

\newcommand{\no}{\noindent}

\newcommand{\tth}{\tilde{h}}

\newcommand{\s}{\,\,\,\,}
\def\bea{\begin{eqnarray}}
\def\eea{\end{eqnarray}}

\newcommand{\beq}{\begin{eqnarray}}
\newcommand{\eeq}{\end{eqnarray}}
\begin{document}

\title{\textbf{Dimensional reduction of the massless limit of the linearized ``New Massive Gravity''}}
\author{H. A. Biazotti\footnote{biazotti@gmail.com}, D. Dalmazi\footnote{dalmazi@feg.unesp.br} and
G. B. de Gracia\footnote{gb9950@gmail.com} \\
\textit{{UNESP - Campus de Guaratinguet\'a - DFQ} }\\
\textit{{Avenida Doutor Ariberto Pereira da Cunha, 333} }\\
\textit{{CEP 12516-410 - Guaratinguet\'a - SP - Brazil.} }\\}
\date{\today}
\maketitle

\begin{abstract}

The so called ``New Massive Gravity'' in $D=2+1$ consists of the Einstein-Hilbert action (with minus sign) plus
a quadratic term in curvatures ($K$-term). Here we perform the Kaluza-Klein dimensional reduction of the
linearized $K$-term to $D=1+1$. We end up with a fourth-order massive electrodynamics in $D=1+1$ described by a
rank-2 tensor. Remarkably, there appears a local symmetry in $D=1+1$ which persists even after gauging away the
Stueckelberg fields of the dimensional reduction. It plays the role of a $U(1)$ gauge symmetry. Although of
higher-order in derivatives, the new $2D$ massive electrodynamics is ghost free, as we show here. It is shown,
via master action, to be dual to the Maxwell-Proca theory with a scalar Stueckelberg field.
%
\end{abstract}

\newpage

\section{Introduction}

The authors of \cite{bht} have suggested an invariant theory under
general coordinate transformations which describes a massive
spin-2 particle (graviton) in $D=2+1$. The model contains the
Einstein-Hilbert theory and an extra term of fourth-order in
derivatives, quadratic in curvatures, the so called $K$-term which
has been analyzed in \cite{deserprl}, see also \cite{unitary}.
 Since massless particles in $D$ dimensions have
the same number of degrees of freedom  of massive particles in
$D-1$ dimensions, one might wonder whether the ``New Massive
Gravity'' (NMG) theory might be regarded as a dimensional
reduction of some fourth-order (in derivatives) massless spin-2
model in $D=3+1$ which would be certainly interesting from the
point of view of a renormalizable quantum gravity in $D=3+1$. As
far as we know there is no positive answer to that question so
far.\footnote{See however \cite{jm} which shows that a
Kaluza-Klein dimensional reduction of the usual ({\bf
second-order}) massless Fierz-Pauli theory followed by an
unconventional elimination of fields and a dualization procedure
leads to the linearized NMG theory.} As an attempt to gain more
insight on that question we investigate here the dimensional
reduction of the massless part of the linearized NMG theory, the
linearized $K$-term. We show here that the linearized $K$-term is
reduced to a kind of higher-derivative massive $2D$
electrodynamics, which is in agreement with the fact that the
linearized $K$-term is dual to the Maxwell theory in $3D$ as shown
in \cite{bht2}, see also \cite{deserprl} and \cite{renato}.
However, it is remarkable that a new local symmetry shows up after
dimensional reduction and plays the role of a $U(1)$ symmetry not
broken by the mass term. We also derive in section 4 a master
action interpolating between the new (higher-order) massive $2D$
electrodynamics and the usual Maxwell-Proca theory with a
Stueckelberg field. We emphasize that throughout this work we only
deal with quadratic (linearized) free theories.

\section{From $2+1$ to $1+1$}

\no Here we take capital indices in three dimensions ($M,N=0,1,2$)
and greek small indices in two dimensions $(\mu,\nu = 0,1)$,
except in the appendix. Expanding about a flat background,
$g_{MN}= \eta_{MN} + h_{MN}$, where $\eta_{MN}=(-,+,+)$, the
$K$-term \cite{bht,deserprl} becomes, in the quadratic
approximation

\bea S_K &=&  \int d^3 \, x  \sqrt{-g}\left( R_{MN} R^{MN} - \frac 38 R^2 \right)_{hh} \label{rr} \\ &=& \frac
14\int d^3 \, x \left\lbrack \left(\Box \theta_{AN}h^{NM} \right)\left(\Box \theta_{MB}h^{BA}\right) -
\frac{\left(\Box\theta_{MN}h^{MN}\right)^2}2 \right\rbrack
\label{skm} \\
&=& \frac 14\int d^3 \, x h^{AB}\left(\Box^2 P_{TT}^{(2)} \right)_{ABCD}h^{CD} \label{sk}\eea

\no where we have the spin-1 projection operator

\be \theta_{MN} = \eta_{MN} - \frac{\p_M\p_N}{\Box} \quad , \ee

\no while $P_{TT}^{(2)}$ is the spin-2 projection operator acting
on symmetric rank-2 tensors  in $D=3$.  It is given in formula
(\ref{ps2}) of the appendix for arbitrary dimensions.

Since projection operators of different spins are orthogonal, it is clear from (\ref{sk}) that there is a
general spin-1 plus a general spin-0 local symmetry in the quadratic approximation for the $K$-term, i.e.,

\be \delta h_{AB} = \p_A \xi_B + \p_B\xi_A + \eta_{AB} \Lambda
\label{sym3d} \ee

\no The vector symmetry corresponds to linearized
reparametrizations as expected from the general covariant form of
the nonlinear theory (\ref{rr}). The Weyl symmetry is surprising
from the point of view of the nonlinear version of (\ref{rr})
since it does not hold beyond the quadratic approximation. It
leads to an awkward situation for perturbation theory, where a
scalar degree of freedom is present in interacting vertices but it
does not propagate, see comments in \cite{deserprl,deg}.
Indirectly the Weyl symmetry leads to an unexpected symmetry in
the reduced theory as we show here.

 In
order to proceed with the dimensional reduction we consider the
second space dimension $(x_2 \equiv y )$ constrained to a circle
of radius $R=1/m$. Explicitly, the action $S_K$ becomes

\be S_K = \frac 14 \int_{0}^{2\pi R} \!\!\! dy \int \, d^2 x \left\lbrack h_{AB}\Box^2 h^{AB} + 2 \p_Ah^{AB}\Box
\p^{C}h_{CB} + \frac 12 (\p_A\p_Bh^{AB})^2 - \frac 12 h \Box^2 h + \p_A\p_Bh^{AB}\Box h \right\rbrack \\
\label{sk2} \ee

\no As usual for Kaluza-Klein reductions, we expand tensor fields
with even (odd) number of indices in the y-dimension in terms of
periodic even (odd) functions, see e.g. \cite{kmu}. Using

\be h_{\mu\nu}(x,y) = \sqrt{\frac{m}{\pi}} h_{\mu\nu}(x) \cos (m\,y) \quad ; \quad h_{\mu,2} (x,y) =
\sqrt{\frac{m}{\pi}} \phi_{\mu}(x) \sin (m\, y) \quad , \label{ansatz1} \ee

\be h_{22}(x,y) = \sqrt{\frac{m}{\pi}} H (x) \cos (m\,y) \quad , \label{ansatz2} \ee

\no back in the action (\ref{sk2}) we obtain the complicated
action in $D=1+1$:

\bea S_{2D} &=& \frac 14\int d^{2}x\left[ m^{4}\left( h_{\nu\mu}h^{\nu\mu}- h^{2}/2 \right) - 2\,
m^{3}\left(2\partial^{\mu}h_{\mu\lambda}\phi^{\lambda}+
\partial_{\beta}\phi^{\beta}h \right)
 \nn \right.
\\  &+& \left. m^{2}\left(-2h_{\nu\mu}\Box
h^{\nu\mu}-2\phi^{\nu}\Box\phi_{\nu} -
2\partial^{\mu}h_{\nu\mu}\partial_{\beta} h^{\nu\beta}
-2\partial_{\nu}\partial_{\mu}h^{\nu\mu}H + H\Box h
-\partial_{\alpha}\partial_{\beta}h^{\alpha\beta}h+ h\Box h
\right)
 \nn \right.
\\ &+& \left. 2\, m(-\partial^{\mu}\phi_{\mu}\Box H
+ 2\, \partial^{\mu}h_{\nu\mu}\Box \phi^{\nu}+
\partial_{\nu}\partial_{\mu}h^{\nu\mu}\partial_{\alpha}\phi^{\alpha}
+ \partial_{\beta}\phi^{\beta}\Box h) \nn \right.
\\ &+& \left. 2\, \phi^{\nu}\Box^{2}\phi_{\nu}+\frac{1}{2}H\Box^{2}H
+ 2\, \Box\partial^{\mu}h_{\nu\mu}\partial_{\beta}
h^{\nu\beta}+\frac{1}{2}(\partial_{\nu}\partial_{\mu}h^{\nu\mu})^{2}
\nn \right.
\\ &-& \left. \frac{1}{2}h\Box^{2}h - H\Box^{2}h+\partial_{\alpha}\partial_{\beta}h^{\alpha\beta}\Box
h  +
\partial_{\alpha}\partial_{\beta}h^{\alpha\beta}\Box H +
2(\partial_{\mu}\phi^{\mu})\Box(\partial_{\nu}\phi^{\nu})+ h_{\mu\nu}\Box^{2}h^{\mu\nu}\label{s2d}\right] \eea

\no Following the same rationale already mentioned, we expand the
parameters of the $3D$ symmetry (\ref{sym3d}) as follows

\be \xi_{\mu}(x,y) = \sqrt{\frac{m}{\pi}} \xi_{\mu}(x) \, \cos (m\,y) \quad ; \quad \xi_2 (x,y) =
\sqrt{\frac{m}{\pi}} \Omega (x) \, \sin (m\, y) \label{ansatz3} \ee

\be \Lambda (x,y) = \sqrt{\frac{m}{\pi}} \Lambda (x) \, \cos (m\,y) \label{ansatz4} \ee

\no Back in (\ref{sym3d}) we deduce the $2D$ symmetry transformations which leave the reduced action (\ref{s2d})
invariant, as we have explicitly checked,

\bea \delta h_{\mu\nu} &=& \p_{\mu}\xi_{\nu} + \p_{\nu} \xi_{\mu} + \eta_{\mu\nu} \Lambda \label{deltah} \\
\delta \phi_{\mu} &=& - m \, \xi_{\mu} + \p_{\mu} \Omega \label{deltaphi} \\
 \delta H &=& \Lambda + 2\, m \, \Omega
\label{deltaH} \eea

Before we go on, it is certainly welcome to simplify the long
expression for $S_{2D}$ in order to figure out its particle
content. We use as a guide the known dimensional reductions of
Mawell  to Maxwell-Proca (spin-1)  and from the massless to the
massive Fierz-Pauli \cite{fp} theory (spin-2). Recall that in the
first case we have

\bea {\cal L}_{Maxwell}^D &=& \frac 12 A^M \left( \Box \eta_{MN} - \p_{M}\p_{N} \right)A^N \quad , \label{max} \\
{\cal L}_{Proca}^{D-1} &=& \frac 12 \tilde{A}^{\mu} \left\lbrack \left(\Box - m^2\right) \eta_{\mu\nu} -
\p_{\mu}\p_{\nu} \right)\tilde{A}^{\nu} \quad . \label{proca} \eea

\no where $\tilde{A}_{\mu} = A_{\mu} + \p_{\mu}\phi/m $ is the only local combination involving the vector field
$A_{\mu}$ and the Stueckelberg scalar $\phi$ (stems from $A_{D-1}$) which is invariant under the reduced gauge
symmetry: $\delta A_{\mu} = \p_{\mu} \epsilon $ ; $ \delta \phi = - m\, \epsilon $. Analogously, in the spin-2
case we have the massless and massive Fierz-Pauli theories respectively,

\bea {\cal L}^D_{FP(m=0)} &=& \frac 14 \left\lbrack h^{MN} \, \Box \, h_{MN} - h \, \Box \, h \right\rbrack +
\frac 12 \p^{A}h_{AM}\left(\p_B h^{BM} - \p^M h \right) \label{eh} \\
{\cal L}^{D-1}_{FP(m\ne o)} &=& \frac 14 \left\lbrack \tth^{\mu\nu} \, \left( \Box -m^2\right) \, \tth_{\mu\nu}
- \tth \, \left( \Box - m^2 \right) \, \tth \right\rbrack + \frac 12 \p^{\alpha}\tth_{\alpha\mu}\left(\p_{\beta}
\tth^{\beta\mu} - \p^{\mu} \tth \right) \label{ehfp}  \eea

\no where $\tth_{\mu\nu} = h_{\mu\nu} + (\p_{\mu}\phi_{\nu} + \p_{\nu} \phi_{\mu})/m - \p_{\mu}\p_{\nu}H/m^2 $
is the only local combination of those fields invariant under the reduced reparametrization symmetry given in
(\ref{deltah})-(\ref{deltaH}) with $\Lambda =0$.

Comparing (\ref{proca}) and  (\ref{max}) for spin-1 and (\ref{ehfp}) with (\ref{eh}) for spin-2, one infers the
rather simple rule, see \cite{ady} which includes spin-3,  for the Kaluza-Klein dimensional reduction:

\bea \Box_D \to \Box_{D-1} - m^2 \quad &;&  \quad A_{M} \to
\tilde{A}_{\mu} = A_{\mu} + \p_{\mu}\phi /m \label{rule1a} \\
\Box_D \to \Box_{D-1} - m^2 \quad &;& \quad h_{MN} \to
\tth_{\mu\nu} = h_{\mu\nu} + (\p_{\mu}\phi_{\nu} + \p_{\nu}
\phi_{\mu})/m - \p_{\mu}\p_{\nu}H/m^2 \label{rule1b} \eea

\no This suggests that the long expression (\ref{s2d}) might be related with (\ref{skm}) via

\be \Box \theta_{MN} \to (\Box - m^2)\eta_{\mu\nu} -
\p_{\mu}\p_{\nu}  \quad ; \quad h_{MN} \to \tth_{\mu\nu} \quad .
\label{rule2} \ee

\no Where $\tth_{\mu\nu}$ is some Stueckelberg combination invariant under (\ref{deltah})-(\ref{deltaH}).  It
turns out that there is no linear combination of the fields $h_{\mu\nu}$ , $\, \left(\p_{\mu}\phi_{\nu} +
\p_{\nu}\phi_{\mu}\right)/m$ , $\,\eta_{\mu\nu} \p \cdot \phi/m$ , $\,\p_{\mu}\p_{\nu}H/m^2$ and  $\,
\eta_{\mu\nu}\, H$ invariant under (\ref{deltah})-(\ref{deltaH}) if $\Lambda \ne 0$. The best we can do is to
stick to $\tth_{\mu\nu}$ as given in (\ref{rule1b}). Under the $2D$ symmetries (\ref{deltah})-(\ref{deltaH}) we
have $\delta \tth_{\mu\nu} = \left( \eta_{\mu\nu} - \frac{\p_{\mu}\p_{\nu}}{m^2} \right) \Lambda $.

Confirming our expectations and following the rules (\ref{rule2}),
the $2D$ action (\ref{s2d}), including the Stueckelberg fields,
can be rewritten in a rather simple way, compare with (\ref{skm}),

\be S_{2D}[\tilde{h}_{\mu\nu}] = \frac 14 \int d^2 \, x \left\lbrack \left( K_{\alpha\nu}\tth^{\nu\mu}
\right)\left(K_{\mu\beta}\tth^{\beta\alpha}\right) - \frac{\left(K_{\mu\nu}\tth^{\mu\nu}\right)^2}2
\right\rbrack \label{s2dk} \ee

\no where

\be K_{\mu\nu} = (\Box - m^2)\eta_{\mu\nu} - \p_{\mu}\p_{\nu}  \label{k} \ee

\no It is now easy to check that (\ref{s2dk}) is invariant under
$\delta \tth_{\mu\nu} = K^{-1}_{\mu\nu} \Lambda $ which becomes
exactly $\delta \tth_{\mu\nu} = \left( \eta_{\mu\nu} -
\frac{\p_{\mu}\p_{\nu}}{m^2} \right) \Lambda $ after we take
$\Lambda \to (\Box - m^2) \Lambda$. Note that the $2D$ symmetries
(\ref{deltah})-(\ref{deltaH}) allow us to gauge away the
Stueckelberg fields: $\phi_{\mu} = 0 = H$ such that
$S_{2D}[\tilde{h}_{\mu\nu}] \to S_{2D}[h_{\mu\nu}]$ which is still
invariant under $\delta h_{\mu\nu} = \left( \eta_{\mu\nu} -
\p_{\mu}\p_{\nu}/m^2 \right) \Lambda $. This is rather surprising
since local symmetries in dimensionally reduced massive theories
usually disappear altogether with the Stueckelberg fields. In
other words, the action $S_{2D}[h_{\mu\nu}]$ is invariant under

\be \delta h_{\mu\nu} = \left( \eta_{\mu\nu} -
\p_{\mu}\p_{\nu}/m^2 \right)\Lambda \quad . \label{nsym} \ee

\no The above local symmetry seems to be technically related with the absence of a linear combination of the
tensors $h_{\mu\nu}$ , $\, \left(\p_{\mu}\phi_{\nu} + \p_{\nu}\phi_{\mu}\right)/m$ , $\,\eta_{\mu\nu} \p \cdot
\phi/m$ , $\,\p_{\mu}\p_{\nu}H/m^2$ and $\, \eta_{\mu\nu}\, H$  invariant under (\ref{deltah}),(\ref{deltaphi})
and (\ref{deltaH}) with $\Lambda \ne 0$.

\section{Gauge invariant massive electrodynamics in $D=1+1$}

\subsection{Equations of motion}

After gauging away the Stueckelberg fields, the fourth-order
equations of motion of $S_{2D}[h_{\mu\nu}]$ are given by

\be (\Box- m^2)\left\lbrack \p_{\mu}V_{\nu} + \p_{\nu} V_{\mu}
-\eta_{\mu\nu}\frac{\p \cdot V}2 - \frac{\p_{\mu}\p_{\nu}h}2 -
(\Box - m^2)(h_{\mu\nu} - \eta_{\mu\nu}\frac h2 ) \right\rbrack =
\frac{\p_{\mu}\p_{\nu} (\p \cdot V )}2 \nn \\ \label{em1}\ee

\no where we have defined the vector field

\be V_{\mu} \equiv \p^{\nu}h_{\mu\nu} \quad. \label{v} \ee

\no From the trace of (\ref{em1}) we have

\be (2\, m^2 - \Box )\p^{\mu}\p^{\nu}h_{\mu\nu} + \Box (\Box -
m^2)h =0 \quad . \label{em2} \ee

\no It is convenient to fix the gauge of the new scalar symmetry
(\ref{nsym}) in a such way that (\ref{em2}) is reduced to
second-order. Namely, we choose the scalar gauge condition

\be K^{\mu\nu} h_{\mu\nu} = \p^{\mu}\p^{\nu}h_{\mu\nu} - (\Box -
m^2)h =0 \quad . \label{gc} \ee

\no From (\ref{em2}) and (\ref{gc}) we have

\bea \p^{\mu}\p^{\nu}h_{\mu\nu} &=& \p \cdot V = 0 \quad ,
\label{div} \\ (\Box - m^2) h &=& 0 \quad . \label{kgh} \eea

\no The gauge condition (\ref{gc})  and all equations so far are
invariant under residual symmetry transformations (\ref{nsym})
with the restriction $(\Box - m^2)\Lambda =0$. In particular, we
have

\be \delta V_{\mu} = (m^2 - \Box )\p_{\mu}\Lambda = 0 \quad .
\label{resv} \ee

\no Under such residual transformations we have $\delta \, h =
(2\, m^2 - \Box)\Lambda = m^2 \Lambda $. Therefore we can use the
residual symmetry to get rid of the trace

\be h =0 \quad . \label{h} \ee

\no From $\p^{\mu}$ on (\ref{em1}) and (\ref{em1}) itself we
deduce

\bea (\Box - m^2)V_{\mu} &=& 0 \quad , \label{kgv} \\
(\Box -m^2)^2 h_{\mu\nu} &=& 0 \quad , \label{kghmn} \eea

\no At this point it is convenient to recall that the particle content of the massive Fierz-Pauli theory in
$D=1+1$ is zero.  It is the same content of the massless FP theory in $D=2+1$. This amounts to say that we have
the trivial identity in $D=1+1$:

\be \Box \left( P_{TT}^{(2)}\right)_{\mu\nu}^{\,\,\, \alpha\beta} h_{\alpha\beta} = \Box h_{\mu\nu} +
(\p_{\mu}\p_{\nu} - \Box \eta_{\mu\nu})h - \p_{\mu}V_{\nu} - \p_{\mu}V_{\nu} + \eta_{\mu\nu}\p \cdot V = 0 \quad
. \label{ident} \ee

\no The reader can check that (\ref{ident}) is not a dynamic
equation and vanishes identically for each of its
components\footnote{The identity (\ref{ident}) corresponds to the
linearized version of the Einstein equation $R_{\mu\nu} =
g_{\mu\nu}R/2$ about a flat background $g_{\mu\nu} = \eta_{\mu\nu}
+ h_{\mu\nu}$. Recall that the Einstein equation is a trivial
identity in $D=1+1$ without any dynamic content.}. From
(\ref{div}),(\ref{h}),(\ref{kgv}) and (\ref{ident}) the equation
(\ref{kghmn}) becomes

\be h_{\mu\nu} = \frac{\p_{\mu}V_{\nu} + \p_{\nu}V_{\mu}}{m^2}
\quad . \label{hv} \ee

 Therefore, we can consider $(\Box - m^2)V_{\mu}=0$ and $\p \cdot V =0$ as our
primary dynamic equations and $V_{\mu}$ as our fundamental vector
field. Thus, we have the same particle content of the
Maxwell-Proca theory as expected from the equivalence of the
linearized $K$-term and the Maxwell theory in $D=2+1$, see
\cite{bht2}, see also \cite{deserprl} and \cite{renato}. In the
next subsection we confirm the particle content of the $2D$ theory
via the analytic structure of the propagator.

\subsection{Propagator and absence of ghosts}

After gauging away ($\tilde{h}_{\mu\nu} \to h_{\mu\nu}$) the
Stueckelberg fields in (\ref{s2dk}) we can define the differential
operator $G^{\mu\nu\alpha\beta}$ via $S_{2D}[h_{\mu\nu}] = \int
d^2x h_{\mu\nu}G^{\mu\nu\alpha\beta}h_{\alpha\beta}$. Due to the
symmetry (\ref{nsym}) we need a gauge fixing term in order to
obtain $G^{-1}$. We can use the same gauge condition (\ref{gc})
and add a gauge fixing term

\be {\cal L}_{GF} = \lambda \left\lbrack
\p^{\mu}\p^{\nu}h_{\mu\nu} + (m^2-\Box)h \right\rbrack^2 \quad .
\label{gf} \ee

\no Suppressing the four indices, the operator $G^{-1}$, in
momentum space, can be written in terms of the operators defined
in the appendix as follows

\bea G^{-1} &=& \frac{4\, P_{TT}^{(2)}}{(k^2+m^2)^2} + \frac{4\, P_{SS}^{(1)}}{m^2(k^2 + m^2)} + \frac{(2
\lambda +
1)}{\lambda (k^2+m^2)^2} P_{TT}^{(0)} \nn \\
&+& \frac{(2 \lambda +1)}{ \lambda \, m^4} P_{WW}^{(0)} + \frac{(1-2\lambda)}{ \lambda \, m^2(k^2 + m^2)}
\left\lbrack P_{TW}^{(0)} + P_{WT}^{(0)}\right\rbrack \label{gmenos1} \eea

\no Notice that although there is no transverse traceless
symmetric tensor in $D=1+1$, see (\ref{ident}), we have included
the operator $ P_{TT}^{(2)}$ in order to bookkeep the massless
poles since $P_{TT}^{(2)}$ is singular at $k^2=0$, we come back to
that point later.

After adding a source term $S_{source} = \int d^2 x \,
h_{\mu\nu}T^{\mu\nu}$ and integrating over the fields $h_{\mu\nu}$
in the path integral we obtain  the two point function saturated
with external sources:

\be {\cal A}_2 (k) = - i \, (T^{\mu\nu})^*(k) \, G^{-1}_{\mu\nu\alpha\beta} T^{\alpha\beta} (k) \quad .
\label{a2k} \ee

\no The particle content of the theory is obtained from the poles
of ${\cal A}_2 (k)$. Because of the symmetry (\ref{nsym}) we need
$\delta S_{source} =0 $ which imposes a constraint on the sources,
in momentum space we have

\be k_{\mu}k_{\nu}T^{\mu\nu} = - \, m^2 T \label{cons} \ee

\no where $T= \eta_{\mu\nu}T^{\mu\nu} = -T_{00} + T_{11} $.

\no Using the constraint (\ref{cons}), suppressing some indices on the left handed side, we obtain

\bea T^* P_{TT}^{(2)} \, T &=& T^*_{\mu\nu} T^{\mu\nu} - \frac{2}{k^2} (k^{\mu}T_{\mu\alpha}^*)\,
(k_{\beta}T^{\beta\alpha}) - \frac{k^2+ 2\, m^2}{k^2}\vert T
\vert^2  \quad , \label{tpt1}\\
 T^* P_{SS}^{(1)} \, T &=& 2 \left\lbrack \frac{(k^{\mu}T_{\mu\alpha}^*)\,
(k_{\beta}T^{\beta\alpha})}{k^2} - \frac{m^4}{k^4} \vert T \vert^2
\right\rbrack  \quad ,  \label{tpt2}\\
T^* P_{TT}^{(0)} \, T &=& \frac{(k^2+m^2)^2}{k^4} \vert T \vert^2 \quad ; \quad T^* P_{WW}^{(0)} \, T =
\frac{m^4}{k^4} \vert T
\vert^2 \quad , \label{tpt3}\\
T^* \left( P_{WT}^{(0)} +  P_{TW}^{(0)} \right) T &=& - \frac{2\, m^2(k^2+m^2)}{k^4} \vert T \vert^2 \quad .
\label{tpt4} \eea

\no Back in (\ref{a2k}) and (\ref{gmenos1})  we end up with

\be {\cal A}_2(k) = - 4\, i \frac{\left\lbrack T_{\mu\nu}^*T^{\mu\nu} + \vert T \vert^2 +
2(k^{\mu}T_{\mu\alpha}^*)\, (k_{\beta}T^{\beta\alpha})/m^2 \right\rbrack }{(k^2 + m^2)^2} \label{a2kb}\ee

\no The dependence on the arbitrary gauge parameter $\lambda$ has canceled out, as expected, as well as the
massless pole $k^2=0$ present in the operators $P_{IJ}^{(s)}$. We are left apparently with a dangerous double
pole at $k^2=-m^2$. Double poles indicate ghosts, see comment in \cite{pvn73}. In order to compute the imaginary
part of the residue ($I_{-m^2}$) at $k^2=-m^2$ we introduce the $2D$ vector $k_{\mu}=(m,\epsilon)$ which implies
$k^2 + m^2 = \epsilon^2 $ and take the limit

\be I_{-m^2} = \Im \lim_{\epsilon \to 0} \epsilon^2 \, {\cal A}_2(k) \label{im} \ee

\no So we only need to compute the numerator of ${\cal A}_2(k) $ up to the order $\epsilon^2$. Back in the
constraint (\ref{cons}) we can eliminate $T^{11} = - 2 \, \epsilon \, T^{01}/m + {\cal O}(\epsilon^3) $.
Consequently,

\be \frac{2(k^{\mu}T_{\mu\alpha}^*)\,
(k_{\beta}T^{\beta\alpha})}{m^2} = 2\left\lbrack \left( 1 -
\frac{5\, \epsilon^2}{m^2}\right) \vert T^{01}\vert^2 - \vert
T^{00}\vert^2 - \frac{\epsilon}{m}\left(T_{00}^* T^{01} -
T^{00}T^*_{01}\right)\right\rbrack  + {\cal O}(\epsilon^3) \quad .
\label{res1} \ee

\be  T_{\mu\nu}^*T^{\mu\nu} + \vert T \vert^2 = 2 \left\lbrack
\left( \frac{4\, \epsilon^2}{m^2} - 1\right)\vert T^{01}\vert^2 +
\vert T^{00}\vert^2 \frac{\epsilon}{m}\left(T_{00}^* T^{01} -
T^{00}T^*_{01}\right)\right\rbrack  + {\cal O}(\epsilon^3) \quad .
\label{res2} \ee

\no Back in (\ref{a2kb}) altogether we have

\be I_{-m^2} = \Im \lim_{\epsilon \to 0} \epsilon^2 \left(\frac{8 \, i \,\epsilon^2 \vert T^{01}\vert^2}{m^2
\epsilon^4} \right) = \frac 8{m^2}\vert T^{01}\vert^2 > 0 \quad . \label{imf} \ee

\no Therefore the particle content of the higher order massive
electrodynamics $S_{2D}$ consists solely of a massive physical
particle, confirming the classical analysis of the previous
subsection. The apparent double pole was in fact a simple pole.
Precisely the same result for $I_{-m^2}$ could have been obtained
by dropping the contribution of the $P_{TT}^{(2)}$ term in
(\ref{gmenos1}). This is in agreement with the fact, already used
in the last subsection, that $P_{TT}^{(2)}$ is identically zero if
$k^2 \ne 0$. Moreover, if we look at (\ref{tpt3}) and (\ref{tpt4})
back in (\ref{gmenos1}) we conclude that the spin-0 operators have
furnished no contribution to the massive pole, which is not
obvious if we only look at (\ref{gmenos1}). So we conclude that
the only relevant contribution stems from the spin-1 operator
$P_{SS}^{(1)}$. The potentially dangerous double pole in the
denominator of $P_{TT}^{(2)}$ in (\ref{gmenos1}) has vanishing
residue due to the absence of massive symmetric, transverse and
traceless rank-2 tensors in $2D$, see (\ref{ident}). This is a
relic of the remarkable analytic structure of the propagator of
the K-term in $D=2+1$, see \cite{unitary}. Although dangerous
poles show up in the K-term and its $2D$ descendant here, their
residues are harmless basically due to the low dimensionality of
the space-time.

\section{Master action and duality}

Since $S_{2D}$ is physically equivalent to a gauge invariant massive electrodynamics, one might wonder whether
there could exist a master action, see \cite{dj}, interpolating between $S_{2D}$ and a Maxwell-Proca theory with
a Stueckelberg scalar field, leading eventually to a dual map between gauge invariants. Indeed, the key point is
to consider a second-order version of the K-term \cite{bht},

\be {\cal L}^{(2)}_K = - \frac 14 \left( f_{MN}f^{MN} - f^2
\right) - f_{MN}G^{MN}(h) \quad . \label{lk2} \ee

\no where $f_{MN} = f_{NM} $ is an auxiliary tensor field and
$G^{MN}(h)$ is the linearized Einstein tensor in $3D$. Integrating
over $f_{MN}$ leads to the K-term given in (\ref{rr}) while
integrating over $h_{MN}$ leads to the pure gauge solution
$f_{MN}=\p_M A_N + \p_N A_M $ whose substitution in (\ref{lk2})
leads to the $3D$ Maxwell theory. Using (\ref{ansatz1}) and
(\ref{ansatz2}) and the decomposition

\be f_{\mu\nu}(x,y) = \sqrt{\frac{m}{\pi}} f_{\mu\nu}(x) \cos
(m\,y) \quad ; \quad f_{\mu,2} (x,y) = \sqrt{\frac{m}{\pi}}
\psi_{\mu}(x) \sin (m\, y) \quad , \label{ansatz5} \ee

\be f_{22}(x,y) = \sqrt{\frac{m}{\pi}} \Phi (x) \cos (m\,y) \quad
, \label{ansatz6} \ee

\no the dimensional reduction of (\ref{lk2}) furnishes

\bea {\cal L}^{(2)}_{2D} =  - \frac 14 (f_{\mu\nu}f^{\mu\nu} - f^2 ) - \frac 12 \psi_{\mu}\psi^{\mu}  + \frac 12
f \, \Phi - f^{\mu\nu}G_{\mu\nu}[\tilde{h},m^2] - 2 \, \psi_{\mu} G^{\mu \, 2} (\tilde{h}) - \Phi \,
G_{22}(\tilde{h}) \quad . \label{l2a} \eea

\no where $\tilde{h}_{\mu\nu}$ is defined in $(\ref{rule1b})$ and

\be G_{\mu \, 2 }(\tilde{h}) = \frac m2 ( \p_{\mu}\tilde{h} - \p^{\alpha}\tilde{h}_{\alpha\mu} ) \quad ; \quad
G_{22}(\tilde{h}) = \frac 12(\Box \tilde{h} - \p^{\alpha}\p^{\beta} \tilde{h}_{\alpha\beta} ) \quad . \ee

\no The quantity $G_{\mu\nu}[\tilde{h},m^2]$  is the linearized $2D$ Einstein tensor after the shift $\Box \to
\Box - m^2$. If we perform the Gaussian integral over $f_{\mu\nu},\psi_{\mu}$ and $\Phi$ we recover
(\ref{s2dk}).

Since the $2D$ Einstein tensor vanishes identically, see
(\ref{ident}), it turns out that $G_{\mu\nu}[h,m^2]=m^2
(h_{\mu\nu} - h \, \eta_{\mu\nu})/2$. Back in (\ref{l2a}) and
Gaussian integrating over $f_{\mu\nu}$ we have the master action

\bea {\cal L}_M &=&  - \frac 12 \psi_{\mu}\psi^{\mu} + m \, \psi^{\mu}(\p^{\alpha}h_{\alpha\mu} - \p_{\mu} h) +
\frac{m^4}4(h_{\mu\nu}^2 - h^2 ) \nn\\
&-& \frac{\Phi^2}2 - \frac{\Phi}2
\left\lbrack\p^{\mu}\p^{\nu}h_{\mu\nu} - (\Box + m^2)h
\right\rbrack + J^{\mu}(\psi_{\mu}/m - \p_{\mu}\Phi /m^2 ) \quad ,
\label{lm} \eea

\no where we have gauged away the Stueckelberg fields
($\tilde{h}_{\mu\nu} \to h_{\mu\nu}$) and  introduced a source
term. The master action is invariant under the $U(1)$ gauge
symmetry

\be \delta \psi_{\mu} = m\, \p_{\mu} \, \varphi \, , \, \delta\Phi
= m^2 \varphi \, , \, \delta h_{\mu\nu} = -\left( \eta_{\mu\nu} -
\p_{\mu}\p_{\nu}/m^2 \right) \varphi \quad . \label{u1} \ee

On one hand, if we integrate over $h_{\mu\nu}$ in the path
integral we derive the Maxwell-Proca theory with a Stueckelberg
field

\be {\cal L}_{MP} = - \frac 1{2\, m^2} \p_{\mu} \psi_{\nu} (
\p^{\mu}\psi^{\nu} - \p^{\nu}\psi^{\mu}) - \frac 12
\left(\psi_{\mu} - \p_{\mu}\Phi /m \right)^2 +
J^{\mu}(\psi_{\mu}/m - \p_{\mu}\Phi /m^2 ) \quad . \label{mp} \ee

On the other hand, if we first integrate over $\psi_{\mu}$ and
$\Phi$ in the master action we have the dual new massive
electrodynamics

\bea {\cal L}_M &=& \frac{\left\lbrack \p^{\mu}\p^{\nu}h_{\mu\nu} - (\Box + m^2)h \right\rbrack^2}8 +
\frac{m^2}2 \left( \p^{\alpha}h_{\alpha\mu} - \p_{\mu}h \right)^2  + \frac{m^4}4 (h_{\mu\nu}^2 - h^2 )
\label{ldual} \\ &+& J^{\mu}B_{\mu}[h] + \frac{\p \cdot J}{2\, m^2} + \frac{J^2}2 \quad . \nn\eea

\no where

\be B_{\mu}[h] = \p^{\alpha}h_{\alpha\mu} + \frac{1}{2\,
m^2}\p_{\mu} \left\lbrack (\Box - m^2)h -
\p^{\alpha}\p^{\beta}h_{\alpha\beta} \right\rbrack \quad .
\label{B} \ee

 The higher derivative  massive electrodynamics in (\ref{ldual})
is exactly the same one given in (\ref{s2d}) as one can check by
use of the identity (\ref{ident}). So we have demonstrated the
duality between the Maxwell-Proca-Stueckelberg theory (\ref{mp})
and $S_{2D}$. Moreover, from derivatives of  (\ref{mp}) and
(\ref{ldual}) with respect to the source $J_{\mu}$ we show that
correlation functions of $B_{\mu}[h]$ in the dual massive
electrodynamics agree, up to contact terms, with correlation
functions of the vector field $\psi_{\mu}/m - \p_{\mu}\Phi /m^2 $
in the Maxwell-Proca-Stueckelberg theory. So we have the dual map

\be B_{\mu}[h] \leftrightarrow \psi_{\mu}/m - \p_{\mu}\Phi /m^2
\quad . \label{dualmap} \ee

\no In particular,  $B_{\mu}$ is invariant under the $U(1)$ transformation (\ref{u1}) just like the right handed
side of (\ref{dualmap}). From the point of view of $U(1)$ transformations the map (\ref{dualmap}) is consistent
with the identification of $\psi_{\mu}$ with $m\left(\p^{\alpha}h_{\alpha\mu} -\p_{\mu} h\right)$ and $\Phi$
with $ \left\lbrack -(\Box + m^2)h + \p^{\alpha}\p^{\beta}h_{\alpha\beta}\right\rbrack /2$. Thus, the three
degrees of freedom $h_{\mu\nu}$ are somehow mapped into the three variables $(\psi_{\mu} , \Phi)$. Moreover, we
notice that the identification of the propagating massive vector field with $\p^{\alpha}h_{\alpha\mu}$ in
subsection 3.1 is consistent with (\ref{B}) and the gauge condition (\ref{gc}).

Now two remarks are in order. First, if we take $\Box_{D-1} \to
\Box_D + m^2 $ in (\ref{proca}), (\ref{ehfp}) and (\ref{s2dk}) we
derive the corresponding massless higher dimensional theories
(\ref{max}),(\ref{eh}) and (\ref{skm}) (for $D=3$) respectively.
However, if we try the same (non-rigorous) inverse dimensional
reduction with the linearized ``New massive gravity'' of
\cite{bht} it turns out that we do not get rid of the $m^2$ in the
corresponding $4D$ theory. Moreover, we have a tachyon at
$k^2=m^2$. Thus, there seems to be no simple Kaluza-Klein
reduction of a fourth-order spin-2 massless model in $4D$ which
might lead to the ``new massive gravity'' of \cite{bht}, see
however the footnote in the introduction. The results of
\cite{rgpty} and \cite{rmt} suggest us that one should try to
obtain \cite{bht} from the dimensional reduction of an extra
discrete dimension, see also \cite{ahcg1,ahcg2}, this is under
investigation. Second, it is worth commenting that, although the
K-term has a nonlinear gravitational completion in $D=2+1$, see
(\ref{rr}), there is no such completion for $S_{2D}$ since there
is no local vector symmetry whatsoever in $S_{2D}$.

\section{Conclusion}

By performing a Kaluza-Klein dimensional reduction of the massless
limit of the linearized ``New Massive Gravity'' (linearized
$K$-term) we have obtained a new $2D$ massive electrodynamics of
fourth-order in derivatives. This is in agreement with the
equivalence of the linearized $K$-term with the Maxwell-theory
\cite{bht2}, see also \cite{deserprl} and \cite{renato}. However,
it is remarkable that the reduced $2D$ theory, although massive,
has local $U(1)$ gauge symmetry even after gauging away the
Stueckelberg fields of the dimensional reduction. The $U(1)$
symmetry (\ref{nsym}) seems to be a consequence of the lack of a
Stueckelberg version of the fundamental field $h_{\mu\nu}$
invariant under both linearized reparametrizations and Weyl
transformations, see comment after (\ref{nsym}).

We have also noticed that the dimensional reduction of the K-term follows the same simple pattern of the usual
spin-1 (Maxwell to Maxwell-Proca) and spin-2 (massless Fierz-Pauli to massive Fierz-Pauli) cases, namely, we
have the practical rule $\Box_D \to \Box_{D-1} - m^2$ altogether with the replacement of the fundamental field
by its Stueckelberg version $h_{MN} \to h_{\mu\nu} + (\p_{\mu}\phi_{\nu} + \p_{\nu} \phi_{\mu})/m -
\p_{\mu}\p_{\nu}H/m^2 $.

We have made a classical and quantum analysis of the particle content of the reduced theory, confirming that,
although of 4th-order in derivatives, it is ghost free and contains only a massive vector field in the spectrum.
In particular, we have found a master action interpolating between the new $2D$ massive electrodynamics and the
Maxwell-Proca theory with a scalar Stueckelberg field and identified a dual map between gauge invariant vector
fields in both theories, see (\ref{dualmap}). A possible non-Abelian extension of the new $2D$ electrodynamics
and the issue of unitarity in the context of the Schwinger mass generation when coupled to fermions are under
investigation.

\section{Acknowledgements}

The work of D.D. is supported by CNPq (304238/2009-0) and FAPESP
(2013/00653-4). The works of H.A.B and G.B.de G. are supported by
CNPq (507064/2010-0). We thank Dr. Karapet Mkrtchyan for reminding
us of \cite{jm}.

\section{Appendix A}

In this appendix we use small Greek indices in $D$-dimensions for
both $D=3$ and $D=2$. Using the spin-0 and spin-1 projection
operators acting on vector fields, respectively,

\be  \omega_{\mu\nu} = \frac{\p_{\mu}\p_{\nu}}{\Box} \quad , \quad
\theta_{\mu\nu} = \eta_{\mu\nu} -
\frac{\p_{\mu}\p_{\nu}}{\Box}\quad , \label{pvectors} \ee

\no as building blocks, one can define the projection and
transition operators in $D$ dimensions acting on symmetric rank-2
tensors,

\be \left( P_{TT}^{(2)} \right)^{\lambda\mu}_{\s\s\alpha\beta} =
\frac 12 \left( \theta_{\s\alpha}^{\lambda}\theta^{\mu}_{\s\beta}
+ \theta_{\s\alpha}^{\mu}\theta^{\lambda}_{\s\beta} \right) -
\frac{\theta^{\lambda\mu} \theta_{\alpha\beta}}{D-1} \quad ,
\label{ps2} \ee

\be \left( P_{SS}^{(1)} \right)^{\lambda\mu}_{\s\s\alpha\beta} =
\frac 12 \left(
\theta_{\s\alpha}^{\lambda}\,\omega^{\mu}_{\s\beta} +
\theta_{\s\alpha}^{\mu}\,\omega^{\lambda}_{\s\beta} +
\theta_{\s\beta}^{\lambda}\,\omega^{\mu}_{\s\alpha} +
\theta_{\s\beta}^{\mu}\,\omega^{\lambda}_{\s\alpha}
 \right) \quad , \label{ps1} \ee

\be \left( P_{TT}^{(0)} \right)^{\lambda\mu}_{\s\s\alpha\beta} =
\frac 1{3} \, \theta^{\lambda\mu}\theta_{\alpha\beta} \quad ,
\quad \left( P_{WW}^{(0)} \right)^{\lambda\mu}_{\s\s\alpha\beta} =
\omega^{\lambda\mu}\omega_{\alpha\beta} \quad , \label{psspww} \ee

\be \left( P_{TW}^{(0)} \right)^{\lambda\mu}_{\s\s\alpha\beta} =
\frac 1{\sqrt{D-1}}\, \theta^{\lambda\mu}\omega_{\alpha\beta}
\quad , \quad  \left( P_{WT}^{(0)}
\right)^{\lambda\mu}_{\s\s\alpha\beta} = \frac 1{\sqrt{D-1}}\,
\omega^{\lambda\mu}\theta_{\alpha\beta} \quad , \label{pswpws} \ee

\no They satisfy the symmetric closure relation

\be \left\lbrack P_{TT}^{(2)} + P_{SS}^{(1)} +  P_{TT}^{(0)} +
P_{WW}^{(0)} \right\rbrack_{\mu\nu\alpha\beta} =
\frac{\eta_{\mu\alpha}\eta_{\nu\beta} +
\eta_{\mu\beta}\eta_{\nu\alpha}}2 \quad . \label{sym} \ee

\end{document}